\documentclass[fleqn,10pt]{wlscirep}
\usepackage[utf8]{inputenc}
\usepackage[T1]{fontenc}
\usepackage{amsmath}
\usepackage{bm}

\title{
Skyrmionic {SU}(6) structured light
}

\author[ ]{Shinichi Saito}
\affil[ ]{Center for Exploratory Research Laboratory, Research \& Development Group, Hitachi, Ltd. Tokyo 185-8601, Japan. }

\affil[ ]{shinichi.saito.qt@hitachi.com}

\usepackage[normalem]{ulem}
\usepackage{xcolor}
\DeclareRobustCommand{\erase}{\bgroup\color{red}\markoverwith{\textcolor{red}{\rule[.5ex]{2pt}{0.5pt}}}\ULon}
\newcommand\temp{\bgroup\color{blue}\markoverwith{\color{blue}{\rule[-0.5ex]{2pt}{0.5pt}}}\ULon}

\begin{abstract}
Skyrmions are topological quasi-particles characterised by local spin textures, which are considered to be robust against structural deformation.
N\'eel and Bloch states are famous examples of skyrmions, which exhibit radical and chiral spin profiles, respectively.
Here, we show a skyrmion can be continuously transformed to an antiskyrmion or various other forms of skymions through underlying symmetry of special unitary group of degree six, SU(6), for photons with spin and orbital angular momentum.
We employ a Lie group theory for coherent photons to describe the SU(6) transformation and establish a relationship between the generator of rotation and its expectation values on a hypersphere of  $\mathbb{S}^{35}$.
We propose a simple experimental setup to control the SU(6) states in combinations with wave-plates and vortex lenses to realise the generalised Euler's formula physically in a Mach-Zehnder interferometer.
We show skyrmionic states are described on higher-order Poincar'e spheres together with a recently proposed skyrmionic torus.
\end{abstract}

\begin{document}
\renewcommand{\hbar}{\mathchar'26\mkern-7.5mu h} 

\flushbottom
\maketitle

\thispagestyle{empty}

\section*{Introduction}

The elementary particle physicist, Tony Skyrme, proposed to solve the mechanism why nucleons such as protons and neutrons are so stable in quantum chromodynamics (QCD) by introducing the concept of the topological invariance \cite{Skyrme61}.
He employed a model Lagrangian to describe interacting meson fields and obtained a classical solution, whose spin texture is characterised by a whirl of spin like a hedgehog, pointing along the radial direction over three-dimensional space  ($\mathbb{R}^3$) \cite{Skyrme61}.
Skyrme conjugated the winding number in the homotopy mapping from three-sphere ($\mathbb{S}^3$) to $\mathbb{S}^3$, described as $\pi_{3}(\mathbb{S}^3)=\mathbb{Z}$, where $\mathbb{Z}$ is the set for all integers, corresponds to the baryon number, $B$, and considered baryons are solitons in the non-linear sigma model \cite{Skyrme62}.
The innovative idea of generating a fermion out of bosons could not attract significant attentions in 1960s, and it took two decades to achieve the canonical quantisation of the topological soliton \cite{Witten79,Adkins83}, which is named a skyrmion \cite{DHoker84}. 
Witten theoretically introduced the $N$ colour charges and developed the theory of the special unitary group, SU($N$), in the the limit of $N \rightarrow \infty$ \cite{Hooft74}, where QCD is equivalent to an effective theory of mesons \cite{Witten79}. 
Remarkably, it was proved that the classical solution of Skyrme becomes exact in this limit \cite{Hooft74} and the quantum fluctuation is suppressed as the higher order effect in $1/N$ expansion, showing a skyrmion is a fermion if $N$ is odd  \cite{Witten79,Adkins83}.
In reality, the number of colour charges is three in QCD, and thus, a skyrmion is indeed quantised as a fermion \cite{Witten79,Adkins83}.
In a modern perspective, however, the standard model describes hadrons based on quarks \cite{Gell-Mann61,Ne'eman61,Gell-Mann64} and SU(6) symmetry \cite{Georgi99,Weinberg05} of a vacuum without introducing skyrmions, which were not discovered as elementary particles unlike quarks.
Consequently, the relationship between a skyrmion \cite{Skyrme61,Skyrme62,Witten79,Adkins83,DHoker84,Hooft74} and SU(6) symmetry \cite{Gell-Mann61,Ne'eman61,Gell-Mann64}  has not been completely established, yet.

Regardless of the absence of a skyrmion in original high-energy elementary particle physics, various forms of skyrmions were  successfully discovered in low-energy condensed-matter physics as quasi-particles in magnets \cite{Doring68,Belavin75,Bogdanov94,Muhlbauer09,Tonomura12,Nagaosa13,Han17,Gobel21},  quantum Hall systems \cite{Ezawa13}, superconductors \cite{Rybakov19,Zhang97}, liquid crystals \cite{Ackerman17}, Bose-Einstein condensates \cite{Ho98}, and photonic systems  \cite{Ranada89,Beckley10,Beckley12,Tsesses18,Gao20,Lin21,Lei21,Zhu24,Shen22b,Parmee22,McWilliam23,Shen23,Cisowski23,Lin24}.
In these systems, skyrmions are also known as baby skyrmions \cite{Belavin75,Piette95}, since they are realised in two-dimensional space ($\mathbb{R}^2$), whose spin texture can be mapped onto the surface of two-sphere ($\mathbb{S}^2$) by a stereographic projection \cite{Penrose84,Han17,Nagaosa13,Gobel21,Shen23}, corresponding to the homotopy mapping of $\pi_{2}(\mathbb{S}^2)=\mathbb{Z}$.
The integer to characterise topology of skyrmions is a skyrmion number, $s$, which is the number of times for spin to wrap over $\mathbb{S}^2$ \cite{Nagaosa13,Gobel21,Shen23}, and the higher order excitation is called a hopfion \cite{Gladikowski97,Faddeev97,Chen13,Kobayashi14,Ackerman17,Rybakov19,Sugic21,Ehrmanntraut23,Wan22,Wang23,Shen23b,Rybakov22}, since the map is related to the Hopf fibration \cite{Hopf31,Penrose84,Urbantke91}.
The simplest Hopf map \cite{Hopf31,Penrose84,Urbantke91} is mathematically defined as a map from $\mathbb{S}^3$ to $\mathbb{S}^2$, $\pi_{3}(\mathbb{S}^2)=\mathbb{Z}$, and the two-level quantum system of spin $1/2$ is an excellent physical realisation of the Hopf map.
In fact, the SU(2) wavefunction \cite{Urbantke91} is described by two complex numbers ($\mathbb{C}^2$), which is normalised to be shown on $\mathbb{S}^3$ with a unit radius, and the wavefunction is mapped to give expectation values of spin, which is shown on $\mathbb{S}^2$, known as the Bloch sphere \cite{Sakurai20}.
It is well-known that the gauge transformation in the global U(1) phase of the wavefunction preserves the spin expectation values, and the trajectory of the wavefunction upon changing the U(1) phase forms a great circle ($\mathbb{S}^1$) on $\mathbb{S}^3$, which is mathematically known as a fibre \cite{Hopf31,Penrose84,Urbantke91}.
Each fibre shown as a circle represents a genuine state with unique expectation values associated with the state, and thus, the circles to give different expectation values will never cross each other, forming Villarceau circles and Clifford parallels upon the Hopf fibration \cite{Hopf31,Penrose84,Urbantke91}. 
By mapping $\mathbb{R}^2$ to fibres of wavefuctions, skyrmions were observed \cite{Doring68,Belavin75,Bogdanov94,Muhlbauer09,Tonomura12,Nagaosa13,Han17,Gobel21,Ezawa13,Ackerman17,Ho98,
Ranada89,Beckley10,Beckley12,Tsesses18,Gao20,Lin21,Lei21,Zhu24,
Shen22b,Parmee22,McWilliam23,Shen23,Cisowski23,Lin24},  
while hopfions were observed \cite{Gladikowski97,Faddeev97,Chen13,Kobayashi14,Ackerman17,Rybakov19,Sugic21,Ehrmanntraut23,Wan22,Wang23,Shen23b,Rybakov22}   
by including the U(1) phase into the map from $\mathbb{R}^3$.
For both skyrmions and hopfions, the Hopf map revealed the topological significance in spin textures to cover the entire Bloch sphere  \cite{Hopf31,Penrose84,Urbantke91,Nagaosa13,Gobel21,Shen23}, and the Hopf map is the correspondence \cite{Hopf31,Penrose84,Urbantke91,Nagaosa13,Gobel21,Shen23} between quantum mechanical wavefunction, represented by $\mathbb{C}^2$, and the classical spin expectation values as observables, represented by $\mathbb{R}^3$.
The Hopf map \cite{Hopf31,Penrose84,Urbantke91,Nagaosa13,Gobel21,Shen23} suggests topological structures are indispensable to understand the crossover from quantum to classical behaviours \cite{Caldeira81,Leggett85,Georgi99,Nakahara03}.

The topological nature of quantum mechanics is already found in the crossover from Ising spin to Heisenberg spin \cite{Stokes51,Poincare92,Glauber63,Sakurai20}.
Ising spin is defined for a classical two-level system, characterised by zero-dimensional sphere, $\mathbb{S}^0=\{ 1, -1 \}$, which is isomorphic to the cyclic group of order two, $\mathbb{Z}_2=\{ 0, 1 \}$, and a basis of classical computing as a bit for gate logic.
On the other hand, Heisenberg spin is defined for a quantum two-level system, characterised by $\mathbb{S}^2$, which is a basis of quantum computing as a qubit \cite{Nielsen00}.
Based on the superposition principle, we can construct a unitary transformation from the spin-up state to the orthogonal spin-down state, whose continuous path can be drawn on $\mathbb{S}^2$, such that a sphere in the higher dimension is a topological structure to show the crossover from classical to quantum states \cite{Stokes51,Poincare92,Glauber63,Sakurai20,Nielsen00}.
Mathematically, this quantum-classical crossover is understood by the extension from $\mathbb{Z}_2$ to SU(2) for describing a state. 
The topological signature of polarisation in light \cite{Stokes51,Poincare92,Glauber63,Max99} was also found in a polarisation ellipse, showing the locus and dynamics of the phase front depending on left and right circularly polarised states and their superposition, which is also described by SU(2) symmetry shown on the Poicar\'e sphere of $\mathbb{S}^2$. 
An important lesson to learn from this primitive example is that we can realise the topological transformation among classically distinguishable states by allowing a quantum superposition state between them in higher dimensions.

Topology and underlying symmetry are also important for structured light, 
which is characterised by an optical vortex with orbital angular momentum (OAM) 
\cite{Coullet89,Allen92,Padgett99,Milione11,Naidoo16,Rosales18,Forbes19,Shen19,Shen20b,Shen21,Angelsky21,Yang21,Andrews21,Forbes21,Ma21,Shen22,Nape23}.
An optical vortex possesses topological charge ($m$), which is a winding number for an optical mode with an azimuthal ($\phi$) dependence of the phase as ${\rm e}^{ i m \phi}$ around the node 
\cite{Coullet89,Allen92,Padgett99,Milione11,Naidoo16,Rosales18,Forbes19,Shen19,Shen20b,Shen21,Angelsky21,Yang21,Andrews21,Forbes21,Ma21,Shen22,Nape23}.
OAM is coming from rotational symmetry of the Laguerre-Gauss mode \cite{Coullet89,Allen92}, which is described by the simplest exponential map of the unitary group of degree one, U(1), i.e., the mapping from $\phi \in \mathbb{R}$ to ${\rm e}^{ i m \phi} \in \mathbb{S}^1$, whose kernel is $\mathbb{Z}$ to ensure $m$ is quantised to be an integer \cite{Fulton04}, such that it corresponds to the isomorphism of U(1) $\cong \mathbb{R}/\mathbb{Z} \cong \mathbb{S}^1$.
Consequently, left ($m=1$) and right ($m=-1$) optical vortices become orthogonal to each other, while the superposition state with certain amplitudes and phases is described by SU(2) symmetry and the associated expectation values of OAM \cite{Padgett99,Milione11,Naidoo16} are shown on the extended Poincar\'e sphere of $\mathbb{S}^2$.
If we utilise both spin and OAM, structured light is known as a vector vortex beam \cite{Rosales18}, which is shown on the higher-order Poincar\'e sphere  \cite{Padgett99,Milione11,Naidoo16,Rosales18,Forbes19,Shen19,Shen20b,Shen21,Angelsky21,Yang21,Andrews21,Forbes21,Ma21,Shen22,Nape23}. 

These significant progresses on manipulating structured light enabled researchers to generate photonic skyrmions  \cite{Ranada89,Beckley10,Beckley12,Tsesses18,Gao20,Lin21,Lei21,Zhu24,Shen22b,Parmee22,McWilliam23,Shen23,Cisowski23,Lin24}.
Among many experimental systems to observe skyrmions \cite{Doring68,Belavin75,Bogdanov94,Muhlbauer09,Tonomura12,Nagaosa13,Han17,Gobel21,Ezawa13,Ackerman17,Ho98,
Ranada89,Beckley10,Beckley12,Tsesses18,Gao20,Lin21,Lei21,Zhu24,
Shen22b,Parmee22,McWilliam23,Shen23,Cisowski23,Lin24}, 
photonics is unique in a sense that skyrmions are realised in a free space, where the system is described by non-interacting free bosons.
In fact, photonic skyrmions are made as a superposition state among states with different spin and OAM, while these basis states are degenerate in energy with the same wavelength to ensure the coherence.
This is a remarkable difference from the original Skyrme model, where a solitary solution with a finite energy emerged from a certain interacting many-body system \cite{Skyrme61,Witten79,Adkins83,DHoker84}. 
SU(6) symmetry of our universe was already spontaneously broken \cite{Nambu59,Georgi99,Weinberg05}, such that it is impossible to examine QCD with exact SU(6) symmetry in the vacuum.
Similarly, Faddeev and Niemi proposed a model Hamiltonian to stabilise a knot using a classical field \cite{Faddeev97}, while Dzyloshinskii-Moriya interaction is considered for magnetic skyrmions  \cite{Doring68,Belavin75,Bogdanov94,Muhlbauer09,Tonomura12,Nagaosa13,Han17,Gobel21}.
These models are important to discuss the energetically favourable ground states and to stabilise skyrmions as low-energy excitations 
\cite{Skyrme61,Witten79,Adkins83,DHoker84,Faddeev97,Doring68,Belavin75,Bogdanov94,Muhlbauer09,Tonomura12,Nagaosa13,Han17,Gobel21}.
On the other hand, non-interacting coherent photons from a laser source are employed for structured light to realise photonic skyrmions \cite{Ranada89,Beckley10,Beckley12,Tsesses18,Gao20,Lin21,Lei21,Zhu24,Shen22b,Parmee22,McWilliam23,Shen23,Cisowski23,Lin24}, and therefore, no energy barrier is expected among various skyrmionic states.
Therefore, photonic skyrmions are ideal to investigate the nature of topological structure associated with the symmetry of coherent photons \cite{Ranada89,Beckley10,Beckley12,Tsesses18,Gao20,Lin21,Lei21,Zhu24,Shen22b,Parmee22,McWilliam23,Shen23,Cisowski23,Lin24}.

The purpose of this wok is to explore the topological stability of photonic skyrmions both theoretically and experimentally.
Here, we show that coherent photons have SU(6) symmetry with variable spin and OAM.
By changing amplitudes and phases for SU(6) basis states, we can continuously transform various skyrmionic states such as N\'eel ("Hedgehog") and Bloch states 
\cite{Ranada89,Beckley10,Beckley12,Tsesses18,Gao20,Lin21,Lei21,Zhu24,Shen22b,Parmee22,McWilliam23,Shen23,Cisowski23,Lin24}.
We also show that skyrmions can be continuously transferred to antiskyrmions and {\it vice versa} through the recently proposed skyrmionic torus \cite{Shen22b}, which is characterised by expectation values of spin and OAM.
The proposed experimental system can be utilised to examine the SU(6) symmetry \cite{Georgi99,Weinberg05} of coherent photons. 

\clearpage
\section*{Results and discussion}
\subsection*{Experimental realisation of SU(6) transformation}

\begin{figure}[h]
\begin{center}
\includegraphics[width=16cm]{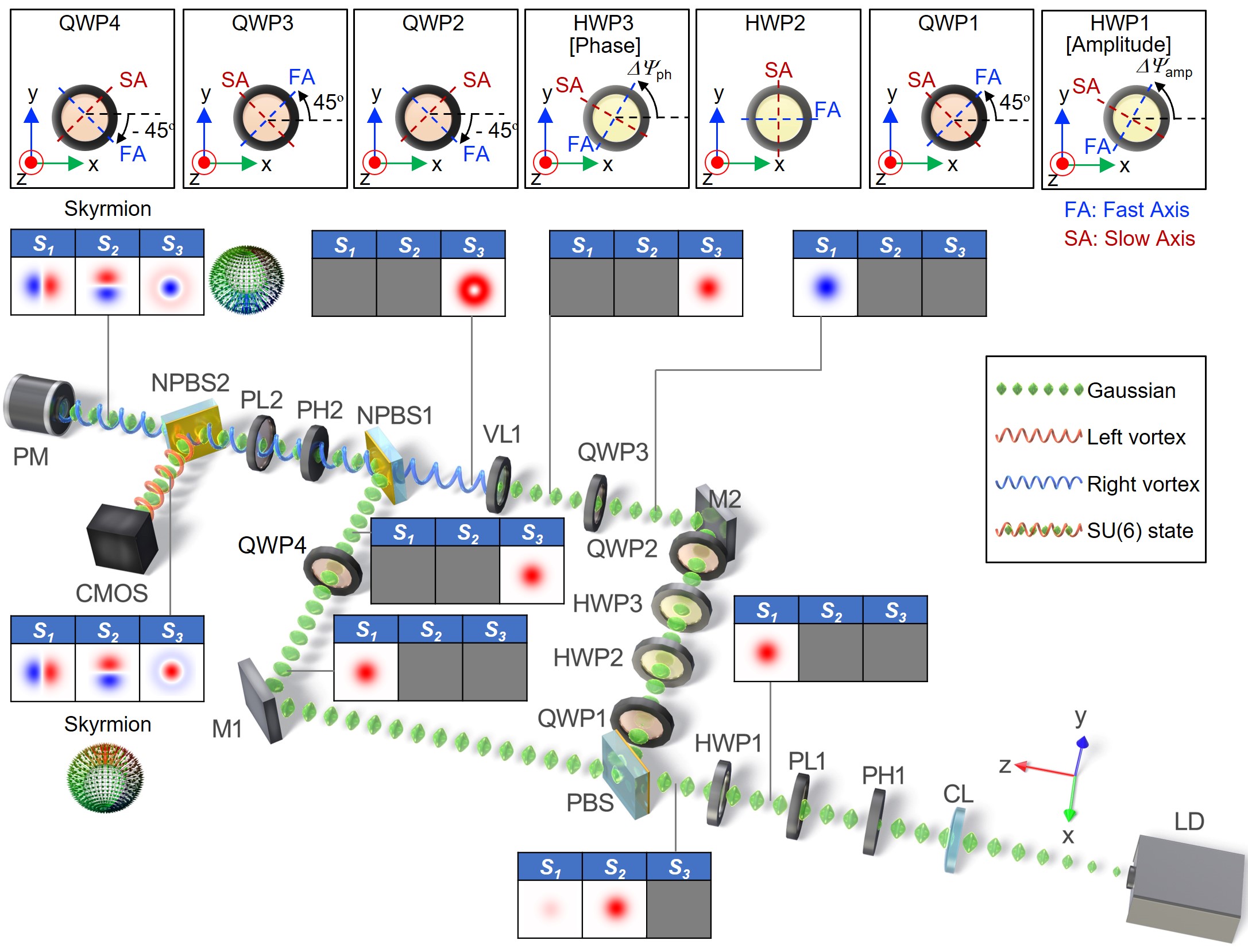}
\caption{
SU(6) transformation for generating skyrmions and antiskyrmions.
The input Gaussian beam is set to be the horizontally (H) polarised state of $|{\rm H} \rangle_{\rm S}|{\rm O} \rangle_{\rm O}$ without an optical vortex.
The half-wave-plate1 (HWP1) works as a rotator, whose fast axis (FA) is rotated with the angle $\Delta \Psi_{\rm amp}$ to control amplitudes for horizontal and vertical (V) components under linear polarisation.
The beam is split into two paths by the polarisation-beam-splitter (PBS); one path going to quarter-wave-plate1 (QWP1), HWP2, HWP3, and QWP2 is dedicated to control the phase without changing $|{\rm V} \rangle_{\rm S}$ by the angle of HWP3, whose FA  is rotated with the angle $\Delta \Psi_{\rm ph}$. 
The beam is then followed by the mirror reflection (M2), polarisation rotation by QWP3, and  vortex generation by the vortex lens (VL1) to make the left circularly polarised ($\uparrow$) right (R) optical vortex, $|\uparrow, {\rm R} \rangle$ for skyrmions; 
the other path going to M1 keeps the H-polarisation until QWP4 rotates it to be the left circular polarised ($\uparrow$)  state without a vortex, $|\uparrow, {\rm O} \rangle$, which is reflected to be $|\downarrow, {\rm O} \rangle$ at the non-polarisation beam splitter (NPBS).
The beam is examined at the polarimeter (PM) for polarisation, and the far-field image is taken at the CMOS camera as a superposition state between $|\uparrow, {\rm O} \rangle = |3 \rangle$ and the right circular polarised left (L) vortex, $|\downarrow, {\rm L} \rangle = |4 \rangle$ for skyrmions.
For antiskyrmions, we just need to flip-flop the VL1 to make a superposition state between $|3 \rangle$ and $|\downarrow, {\rm R} \rangle = |5 \rangle$ with a right vortex at the camera.
}
\end{center}
\end{figure}

The experimental setup is shown in Fig. 1 (Methods).
The primary idea is to realise a generalised Euler's formula, 
\begin{eqnarray}\exp
\left(
-i
\hat{\mathcal S}_{j}
\frac{\delta \phi}{2}
\right) 
&=&
{\bf 1}
\cos
\left(
\frac{\delta \phi}{2}
\right) 
-
i
\hat{\mathcal S}_{j}
\sin 
\left(
\frac{\delta \phi}{2}
\right) , 
\end{eqnarray} 
physically in a Mach-Zehnder interferometer \cite{Saito23k}; we prepare two beams separated at the polarisation-beam-splitter (PBS) with appropriate polarisation controls to change amplitudes, determined by the rotation angle of $\delta \phi/2$; one of the beam is passing through the vortex lens (VL) to generate the beam with OAM, followed by polarisation rotation by the quarter-wave-plate (QWP), corresponding to an $\frak{su(6)}$ operation of $\hat{\mathcal S}_{j}$ to change both spin and OAM simultaneously; and two beams are recombined at the non-polarisation-beam-splitter (NPBS).
There are wide ranges of polarisation components available in a market, such that it is easier to control polarisation rather than controlling an OAM state directly through a dedicated device.
The half-wave-plate (HWP) is equivalent to Pauli matricies of $\hat{\sigma}_1$ and $\hat{\sigma}_2$, depending on the alignment of the fast axis (FA), and the VLs work as ladder operators of $\hat{\lambda}_{+}^{\rm L}$ and $\hat{\lambda}_{+}^{\rm R}$ to generate left and right optical vortices.  
It is also worth noting that chiralities for spin and OAM are reversed upon the mirror reflection at NPBS2, such that we need to consider the final output images to realise skyrmions and antiskyrmions at the CMOS camera.
The mirror reflection cannot convert a skyrmion to an antiskyrmion, but it reverses the direction for wrapping to represent a spin texture.

\subsection*{Poincar\'e spheres for skyrmions and antiskyrmions}

\begin{figure}[h]
\begin{center}
\includegraphics[width=16cm]{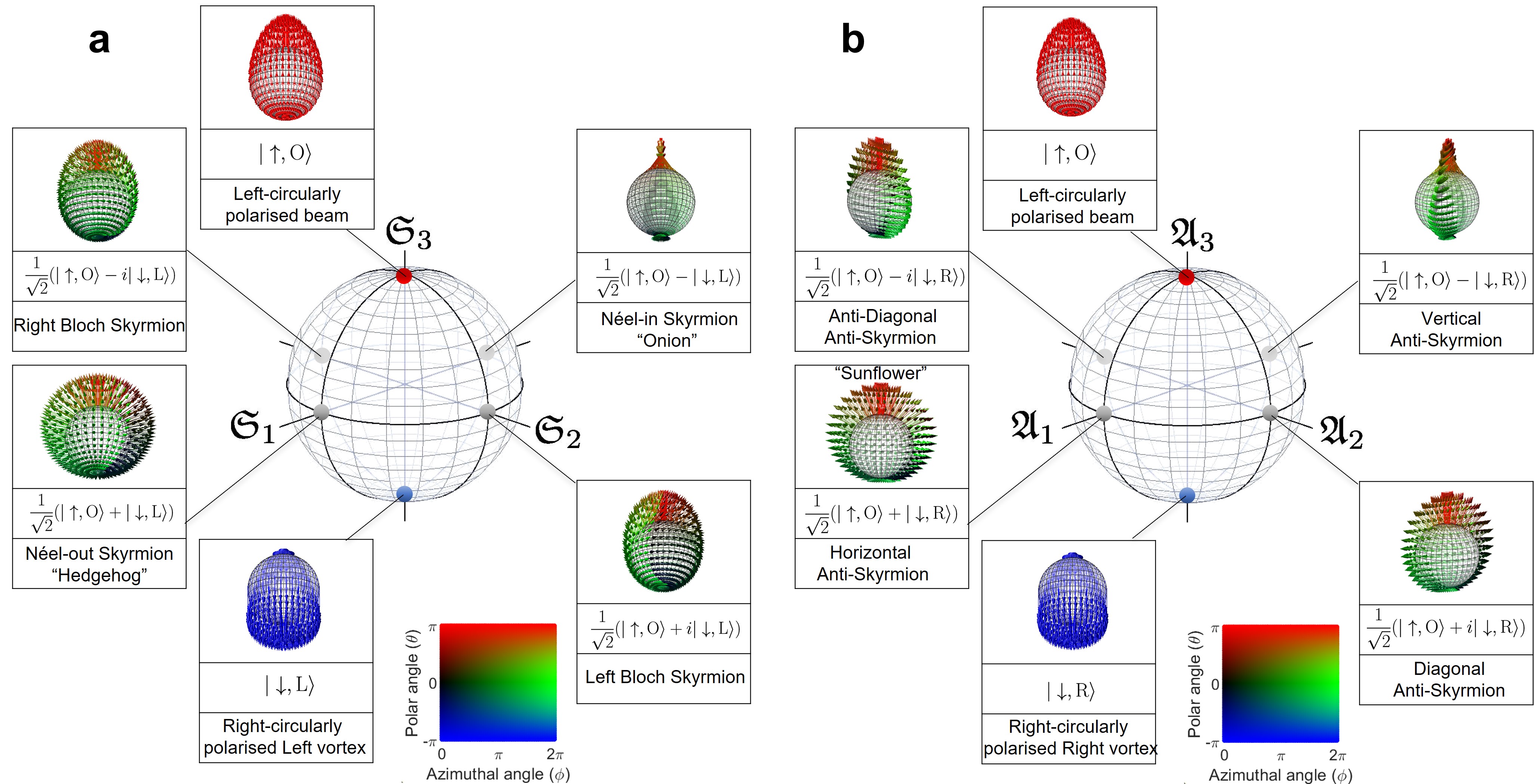}
\caption{
Theoretical expectation for spin textures of photonic skyrmions and antiskyrmions. 
The radius of the sphere is preserved to be $\hbar N_0$ for arbitrary superposition states, made of two orthogonal states of spin and OAM.
{\bf a} Poincar\'e sphere for skyrmions. 
$({\frak S}_1,{\frak S}_2,{\frak S}_3)$ are expectation values of $\frak{su(2)}$ generators of rotations to characterise skyrmions.
The spin textures are changed upon shifting the phase along the equator, from N\'eel-out, left Bloch, N\'eel-in to right Bloch states.
{\bf b} Poincar\'e sphere for antiskyrmions.
$({\frak A}_1,{\frak A}_2,{\frak A}_3)$ are expectation values of $\frak{su(2)}$ generators of rotations to characterise antiskyrmions.
The spin texture is topologically preserved, while the orientation is changed along the equator.
}
\end{center}
\end{figure}

Theoretical expectations of Poincar\'e spheres for photonic skyrmions and antiskyrmions \cite{Ranada89,Beckley10,Beckley12,Tsesses18,Gao20,Lin21,Lei21,Zhu24,Shen22b,Parmee22,McWilliam23,Shen23,Cisowski23,Lin24} 
 are shown in Fig. 2 (Methods).
The most famous skyrmion exhibits a staggered spin texture similar to an anti-ferromagnet and thus, named N\'eel-out skyrmion (Fig. 2a), also known as "Hedgehog", located at 
$({\frak S}_1,{\frak S}_2,{\frak S}_3)=\hbar N_0(1,0,0)$, whose SU(6) state is given by 
$|{\rm N\acute{e}el \ out} \rangle=(|3\rangle + |4 \rangle)/\sqrt{2}$.
If we rotate the state along the equator, we can realise 
the Bloch state with the left chirality, 
$|{\rm Left \ Bloch} \rangle=(|3 \rangle + i |4  \rangle)/\sqrt{2}$ at $\hbar N_0(0,1,0)$, 
the N\'eel-in state like an "Onion", 
$|{\rm N\acute{e}el \ in} \rangle=(|3 \rangle - |4 \rangle)/\sqrt{2}$ at $\hbar N_0(-1,0,0)$, and
the Bloch state with the right chirality, 
$|{\rm Right \ Bloch} \rangle=(|3 \rangle - i |4 \rangle)/\sqrt{2}$ at $\hbar N_0(0,-1,0)$.
The north pole is assigned to the purely left-circularly polarised Gaussian beam of $|3 \rangle$, 
and the south pole corresponds to the left OAM state under right-circular polarisation of $|4 \rangle$.
The states located at the opposite side of the Poincar\'e sphere are orthogonal to each other, 
$\langle {\rm N\acute{e}el \ out} | {\rm N\acute{e}el \ in} \rangle =
\langle {\rm Left \ Bloch} | {\rm Right \ Bloch}  \rangle =0$, because of the orthonormal conditions, 
$\langle 3 | 3 \rangle=\langle 4 | 4 \rangle=1$ and 
$\langle 3 | 4 \rangle=0$, 
similar to polarisation states, $\langle {\rm H}| {\rm V} \rangle=\langle {\rm D}| {\rm A} \rangle=\langle {\rm L}| {\rm R} \rangle = 0$.
 
The antiskyrmions are also shown in Fig. 2b.
The horizontal antiskyrmion is described by 
$|{\rm Antiskyrmion} \rangle=(|3 \rangle + |5 \rangle)/\sqrt{2}$ at 
$({\frak A}_1,{\frak A}_2,{\frak A}_3)=\hbar N_0(1,0,0)$, which looks like a "Sunflower".
Upon changing the phase along the equator, the spin texture of the antiskyrmion is rotated without changing the spin texture. 
Therefore, they are characterised by the orientation of the spin texture, similar to linearly polarised states.
The north pole at $({\frak A}_1,{\frak A}_2,{\frak A}_3)=\hbar N_0(0,0,1)$ is designated for $|3 \rangle$, which also corresponds to $({\frak S}_1,{\frak S}_2,{\frak S}_3)=\hbar N_0(0,0,1)$.
This implies that there is a path to change continuously from a skyrmion to an antiskyrmion, but this is not clear from these Poincar\'e spheres.
On the other hand, the south pole is assigned for the right vortex, $|5 \rangle$ at $({\frak A}_1,{\frak A}_2,{\frak A}_3)=\hbar N_0(0,0,-1)$, which is purely polarised to be right-circular polarisation, such that we cannot distinguish the chirality of OAM from the spin textures of $| 3 \rangle$ and $| 4 \rangle$.

\clearpage
\subsection*{Phase-shifter operation for skyrmions}

\begin{figure}[h]
\begin{center}
\includegraphics[width=16cm]{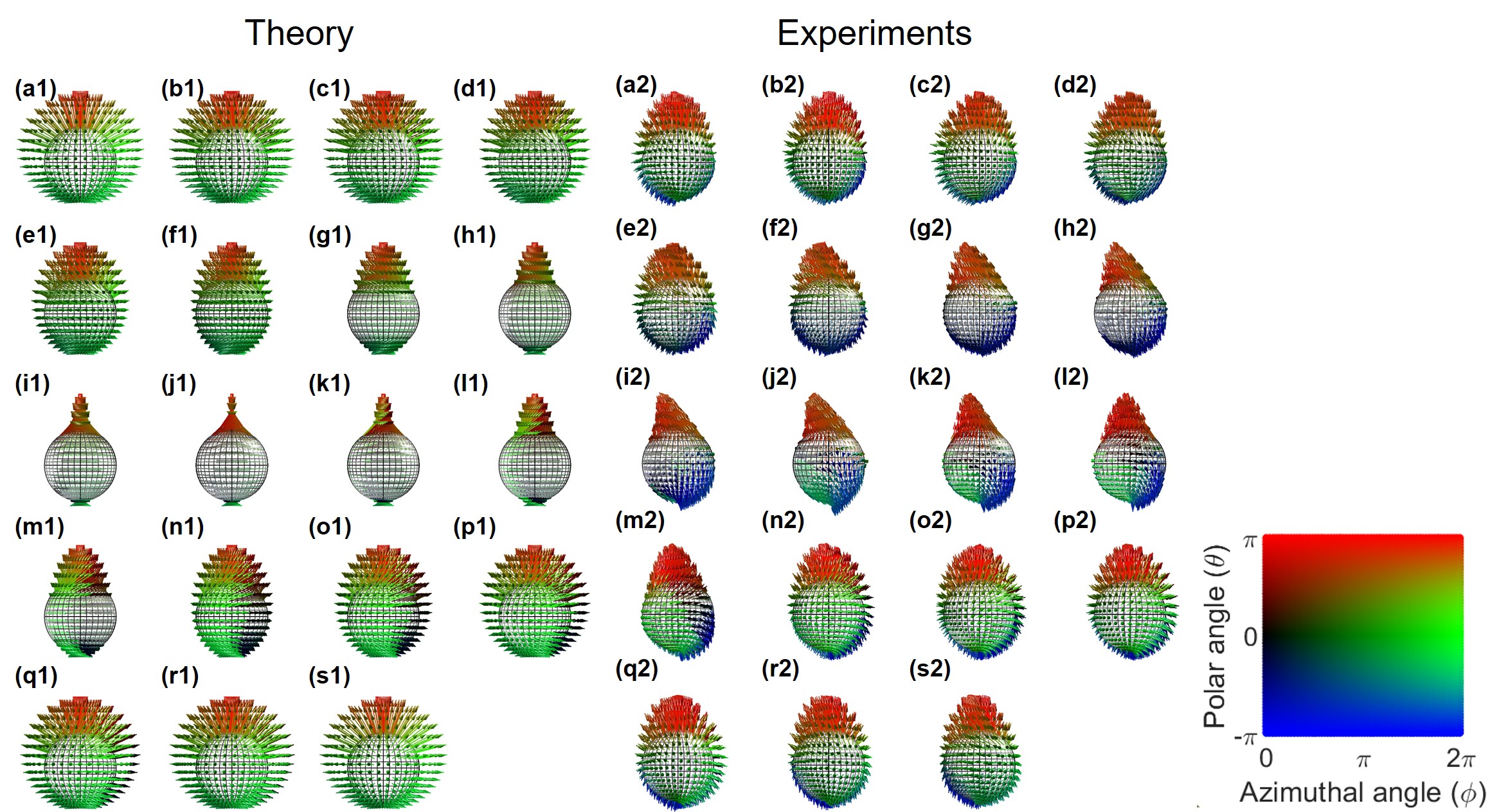}
\caption{
Spin textures of photonic skyrmions upon phase-shifter operations.
Skyrmionic states were rotated on the $\frak{S}_1$-$\frak{S}_2$ plane on the Poincar\'e sphere.
{\bf (a1)-(s1)} theoretical calculations.
{\bf (a2)-(s2)} experiments.
The images were shown upon changing the phase by rotating the HWP3 for the angle of $\Delta \Psi_{\rm ph}$ from $0^{\circ}$ to $180^{\circ}$ with a step of $10^{\circ}$.
(a1) and (a2) N\'eel-out, "Hedgehog"; (f1) and (f2) right Bloch; (j1) and (j2) N\'eel-in, "Onion"; (n1) and (n2) left Bloch states, respectively.
}
\end{center}
\end{figure}

\begin{figure}[h]
\begin{center}
\includegraphics[width=16cm]{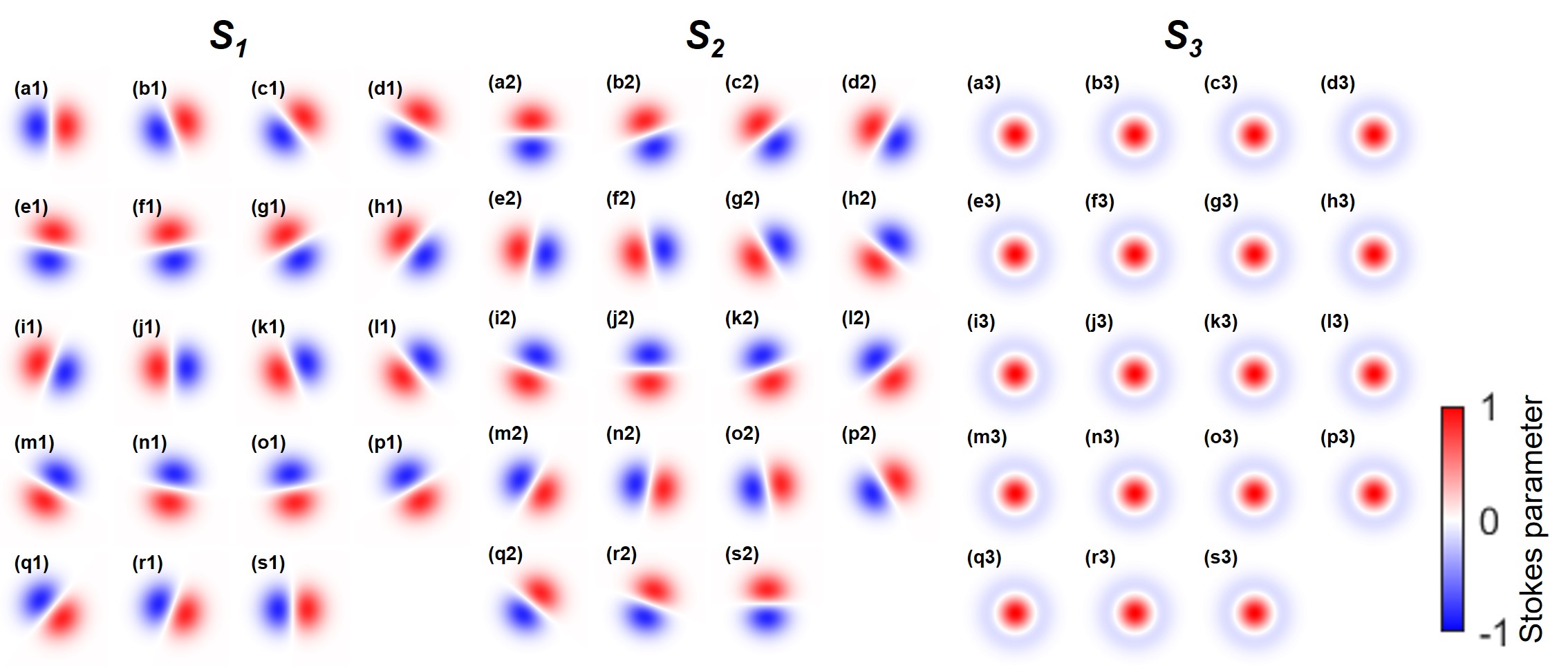}
\caption{
Theoretical local spin profiles for photonic skyrmions upon phase-shifter operations.
Calculated Stokes parameters $(S_1, S_2, S_3)$ are shown.
The images were calculated for the phase change of $\phi_{\rm s}$ from $0^{\circ}$ to $360^{\circ}$ with a step of $20^{\circ}$.
{\bf (a1)-(s1)} $S_1$.
{\bf (a2)-(s2)} $S_2$.
{\bf (a3)-(s3)} $S_3$.
$S_1$ and $S_2$ show photonic dipole after projection of polarisation state to horizontal or diagonal directions, respectively, while $S_3$ is preserved upon changing the phase in the chiral basis.
}
\end{center}
\end{figure}

\begin{figure}[h]
\begin{center}
\includegraphics[width=16cm]{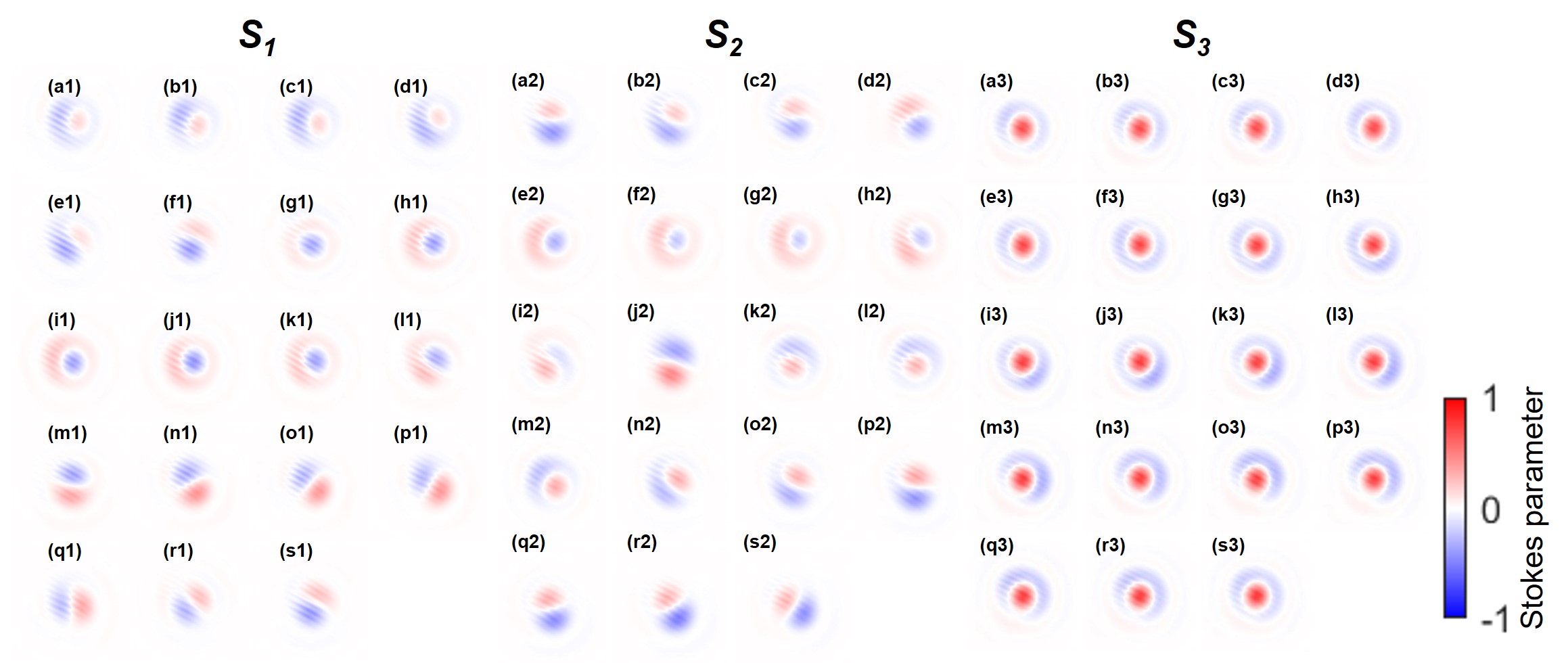}
\caption{
Experimental local spin profiles for photonic skyrmions upon phase-shifter operations.
Stokes parameters $(S_1, S_2, S_3)$ measured from far-field images after passing through waveplates and polarisers are shown.
The images were shown upon changing the phase by rotating the HWP3 for the angle of $\Delta \Psi_{\rm ph}$ from $0^{\circ}$ to $180^{\circ}$ with a step of $10^{\circ}$.
{\bf (a1)-(s1)} $S_1$.
{\bf (a2)-(s2)} $S_2$.
{\bf (a3)-(s3)} $S_3$.
$S_1$ and $S_2$ show rotation of dipoles to the counter-clock-wise direction upon increasing the phase.
}
\end{center}
\end{figure}

We consider rotations over skyrmionic Poincar\'e sphere of Fig. 2a.
It is useful to define an SU(2) state, given by 
$(\alpha_3, \alpha_4)=\sqrt{N_0}({\rm e}^{-i\phi_{\rm s}/2} \cos (\theta_{\rm s}/2) , {\rm e}^{i\phi_{\rm s}/2} \sin (\theta_{\rm s}/2)) \in {\mathbb C}^2$, 
which maps to the point  
$({\frak S}_1,{\frak S}_2,{\frak S}_3)=
\hbar N_0(\sin \theta_{\rm s} \cos \phi_{\rm s} , \sin \theta_{\rm s} \sin \phi_{\rm s}, \cos \theta_{\rm s}) 
\in {\mathbb S}^2$, such that $\theta_{\rm s}$ and $\phi_{\rm s}$ are polar and azimuthal angles, respectively.
We show a phase-shifter operator for skyrmions to change the spin texture along the equator of Fig. 2 on the ${\frak S}_1$-${\frak S}_2$ plane.
This was achieved by changing the phase of $\phi_{\rm s}$ for the state 
$(|3 \rangle + {\rm e}^{i\phi_{\rm s}} |4 \rangle)/\sqrt{2}$. 
In the experiment shown in Fig. 1, the rotation angle of $\Delta \Psi_{\rm ph}$ of HWP3 induced the phase change as $\phi_{\rm s}=-2\Delta \Psi_{\rm ph}$, where the negative sign accounts for the clock-wise rotation in our experiment and the factor of two \cite{Saito23k} is coming from the mirror projection of the polarisation state due to the rotated HWP.

Spin textures upon changing the phase are shown in Fig. 3. 
We used a mapping, named a soup-bubble mapping similar to stereographic mapping, in order to map the local spin distribution taken from a finite far-field image of a closed disk (${\mathbb D}^2$) onto a sphere of  ${\mathbb S}^2$.
We confirmed continuous changes from N\'eel-out (Figs. 3(a1) and 3(a2)), right Bloch (Figs. 3(f1) and 3(f2)), N\'eel-in (Figs. 3(j1) and 3(j2)) to left Bloch (Figs. 3(n1) and 3(n2)) states both theoretically and experimentally.
At each local point in space shown on ${\mathbb S}^2$, local spin was rotated for $360^{\circ}$ in the clock-wise direction, as expected for the change of $\Delta \Psi_{\rm ph}$ for $180^{\circ}$.
The N\'eel-out ("Hedgehog", Figs. 3(a1) and 3(a2)) state is characterised by the local spin pointing outwards from the centre, which was continuously changed to be the opposite state of the N\'eel-in ("Onion", Figs. 3(j1) and 3(j2)) state, whose local spin is pointing inwards to the centre.
The left Bloch (Figs. 3(f1) and 3(f2)) state has the spin texture of left circulation seen from the north pole, while the right Bloch (Figs. 3(f1) and 3(f2)) has the opposite chirality of the right circulation on ${\mathbb S}^2$.

We have also shown theoretical calculations of local spin profiles (Fig. 4) and corresponding experiments (Fig. 5).
They are in reasonable agreement to recognise the topological features, while experimental images were distorted, quantitatively.
The main source of the distortions was coming from the precise alignment, required to combine the beams, and we also estimated the deviation of around $\sim2^{\circ}$ upon mechanical rotations.
Nevertheless, we could confirm the trend that photonic dipole images for $S_1$ (Figs. 4(a1)-4(s1) and 5(a1)-5(s1)) and $S_2$ (Figs. 4(a2)-4(s2) and 5(a2)-5(s2)) rotate to the counter-clock-wise direction upon increasing the phase, while images of $S_3$ (Figs. 4(a3)-4(s3) and 5(a3)-5(s3)) were preserved, because we have not changed the amplitudes among $|3 \rangle$ and $|4\rangle$. 
 
\clearpage

\subsection*{Rotator operation for skyrmions}

\begin{figure}[h]
\begin{center}
\includegraphics[width=16cm]{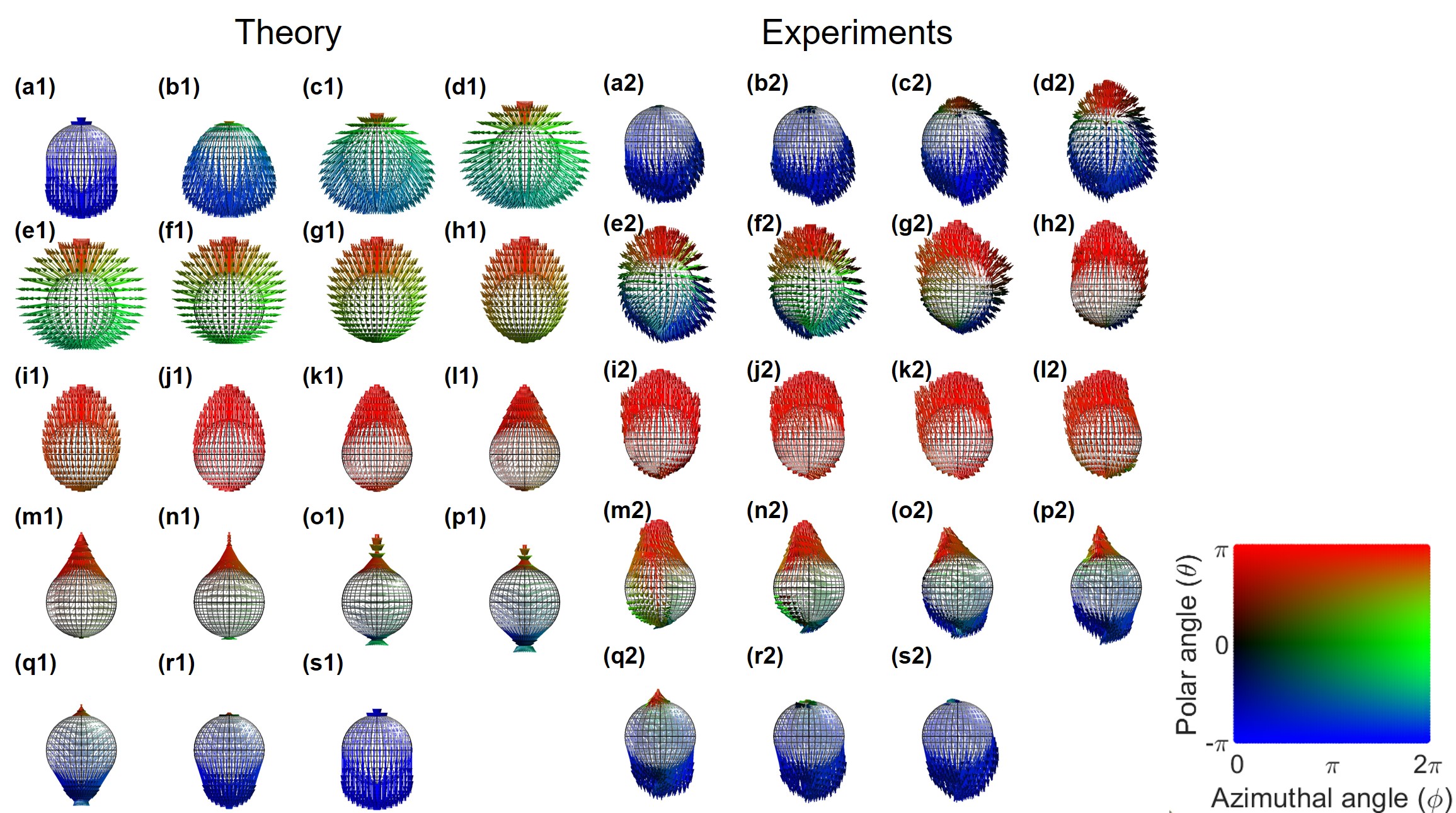}
\caption{
Spin textures of photonic skyrmions upon rotator operations.
Skyrmionic states were rotated on the $\frak{S}_1$-$\frak{S}_3$ plane along $\frak{S}_2$ on the Poincar\'e sphere.
{\bf (a1)-(s1)} theoretical calculations.
{\bf (a2)-(s2)} experiments.
The images were shown upon changing the phase by rotating the HWP1 for the angle of $\Delta \theta_{\rm s}$ from $0^{\circ}$ to $90^{\circ}$ with a step of $5^{\circ}$.
(a1, a2) Right-circularly-polarised left vortex; (e1, e2) N\'eel-out; (j1, j2) left-circularly-polarised Gaussian; (o1, o2) N\'eel-in states, respectively.
}
\end{center}
\end{figure}

\begin{figure}[h]
\begin{center}
\includegraphics[width=16cm]{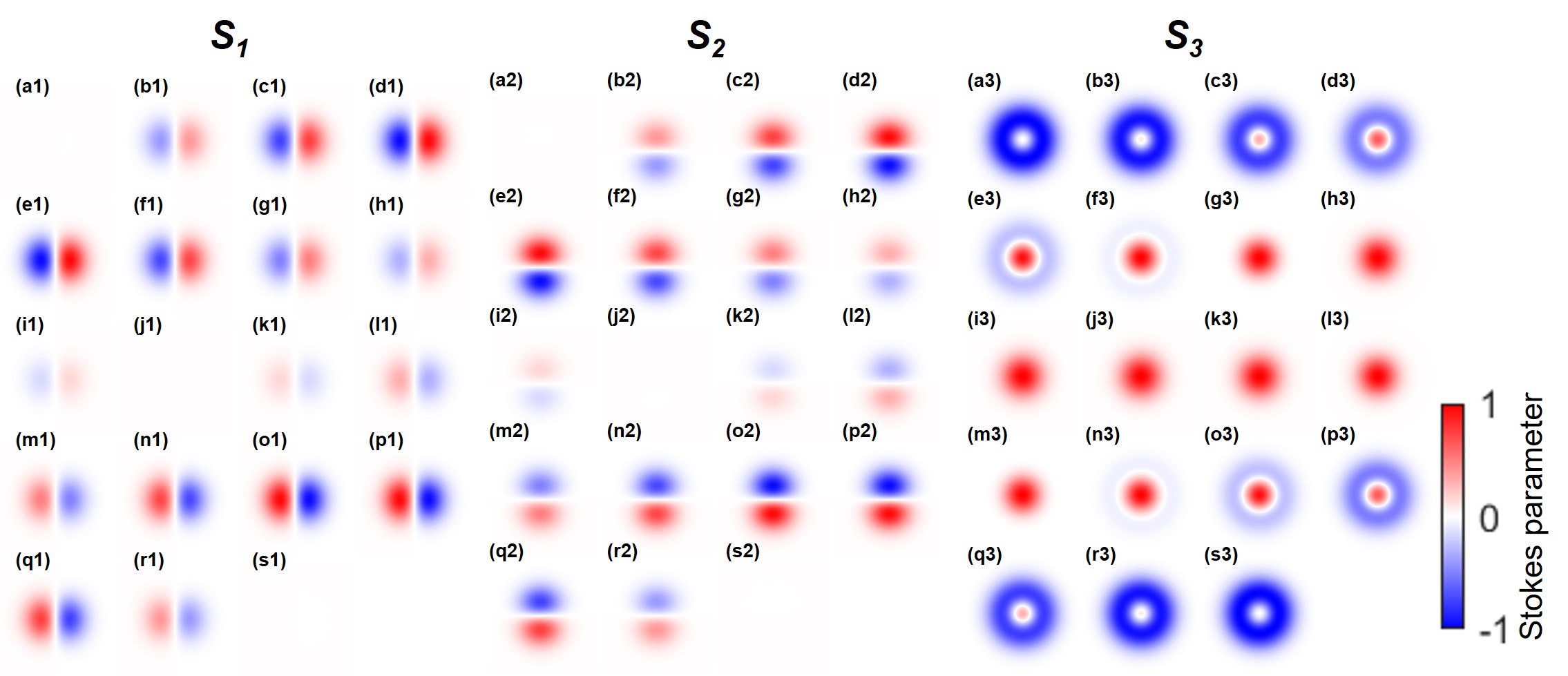}
\caption{
Theoretical local spin profiles for photonic skyrmions upon rotator operations.
Calculated Stokes parameters $(S_1, S_2, S_3)$ are shown.
The images were calculated for the change in the amplitude given by $\theta_{\rm s}$ from $0^{\circ}$ to $360^{\circ}$ with a step of $20^{\circ}$.
{\bf (a1)-(s1)} $S_1$.
{\bf (a2)-(s2)} $S_2$.
{\bf (a3)-(s3)} $S_3$.
$S_1$ and $S_2$ show photonic dipole, if the state is a superposition state between a left vortex and a Gaussian beam, while they disappear, if orbitals are purely made of chiral states at (a1), (a2), (j1), (j2), (s1), and (s2), as is evident in $S_3$ at (a3), (j3), and (s3), respectively. 
N\'eel-out ("Hedgehog") state corresponds to (e1, e2, e3), while the orthogonal state of N\'eel-in ("Onion") corresponds to (o1, o2, o3), which is characterised by the opposite spin direction in the $(S_1,S_2)$ plane, while $S_3$ is the same.
}
\end{center}
\end{figure}

\begin{figure}[h]
\begin{center}
\includegraphics[width=16cm]{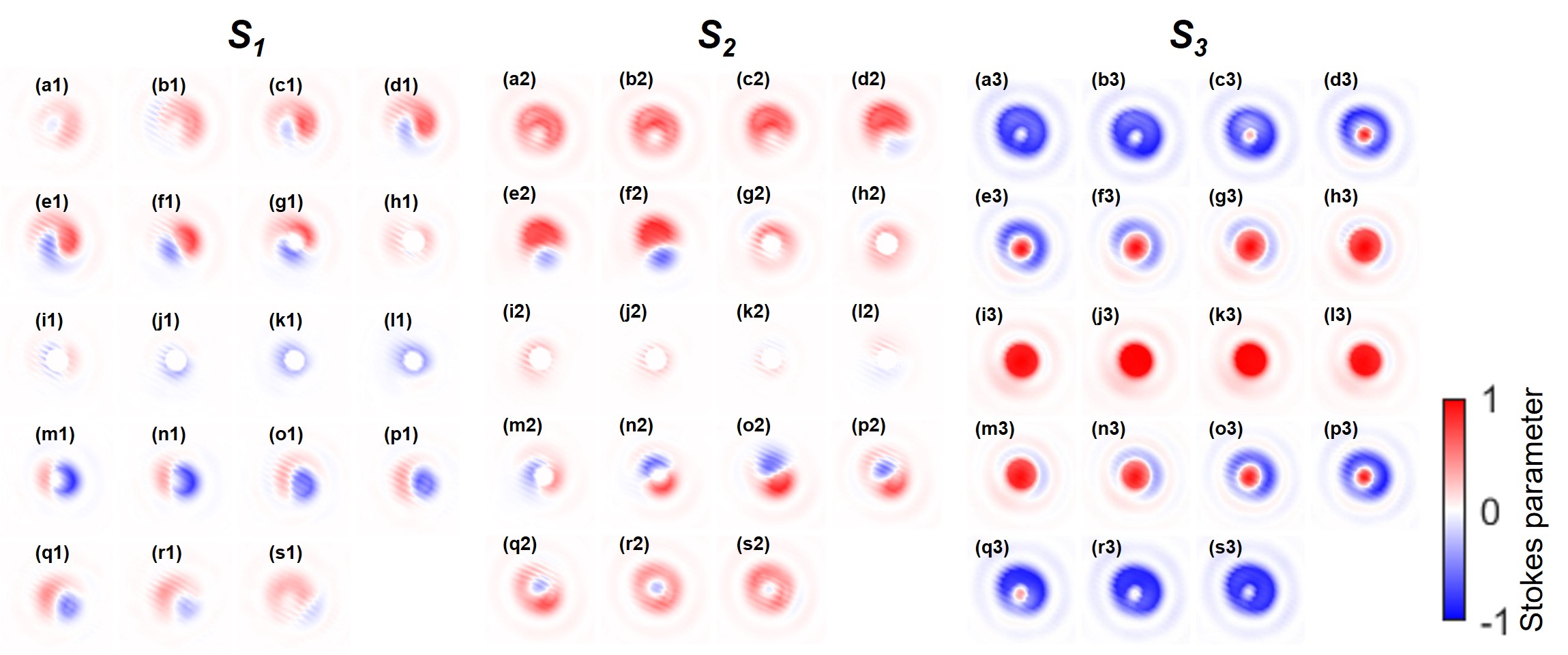}
\caption{
Experimental local spin profiles for photonic skyrmions upon rotator operations.
Stokes parameters $(S_1, S_2, S_3)$ measured from far-field images after passing through waveplates and polarisers are shown.
The images were shown upon changing the phase by rotating the HWP1 for the angle of $\Delta \Psi_{\rm amp}$ from $0^{\circ}$ to $90^{\circ}$ with a step of $5^{\circ}$.
{\bf (a1)-(s1)} $S_1$.
{\bf (a2)-(s2)} $S_2$.
{\bf (a3)-(s3)} $S_3$.
The directions of spin become opposite betwen N\'eel-out (e1, e2) and N\'eel-in (o1, o2) states.
}
\end{center}
\end{figure}

Next, we have examined a rotator operation for skyrmions to change the amplitude of $\theta_{\rm s}$ by rotating HWP1 for the amount of $\Delta \Psi_{\rm amp}$ along the clock-wise direction to give $\theta_{\rm s}=-4\Delta \Psi_{\rm amp}$, which is twice compared with that for the phase-shifter, since both horizontal and vertical components were rotated by HPW1, while only the vertical component was affected by HWP3 \cite{Saito23k}.
HWP1 allowed to change the splitting ratio to change the amplitudes for $|3\rangle $ and $4 \rangle$.
We have adjusted HPW3 to realise the rotation along the $\frak{S}_2$ to rotate in the $\frak{S}_1$-$\frak{S}_3$ plane (Fig. 6).
We confirmed the expected rotation from right-circularly-polarised left vortex (Figs. 6(a1) and 6(a2)), N\'eel-out (Figs. 6(e1) and 6(e2)), left-circularly-polarised Gaussian (Figs. 6(j1) and 6(j2)) to N\'eel-in (Figs. 6(o1) and 6(o2)) states.

The local spin profiles are also calculated as shown in Fig. 7, which was experimentally confirmed in Fig. 8.
The pure OAM state of $|4 \rangle$ has no component on $\frak{S}_1$ (Fig. 7(a1) and Fig. 8(a1)) and $\frak{S}_2$ (Fig. 7(a2) and Fig. 8(a2)) due to the right-circular-polarisation (Fig. 7(a3) and Fig. 8(a3)), and it was gradually converted to the N\'eel-out skyrmion (Figs. 7(e1, e2, e3) and Fig. 8(e1, e2, e3)), followed by the pure Gaussian state with left-circular-polarisation (Figs. 7(j1, j2, j3) and Fig. 8(j1, j2, j3)).
We also confirmed that directions of spin components of $S_1$ and $S_2$ become opposite to each other between N\'eel-out (Figs. 7(e1, e2) and Figs. 8(e1, e2)) and N\'eel-in (Figs. 7(o1, o2) and Figs. 8(o1, o2)) skyrmions.
These results prove that we have a continuous path to transform from a standard Gaussian beam without vortex to a photonic skyrmion or {\it vice versa}.
We have also confirmed that various spin textures of skyrmions can be mutually transferred.

\clearpage

\subsection*{Topological transformation from skyrmions to antiskyrmions}

\begin{figure}[h]
\begin{center}
\includegraphics[width=16cm]{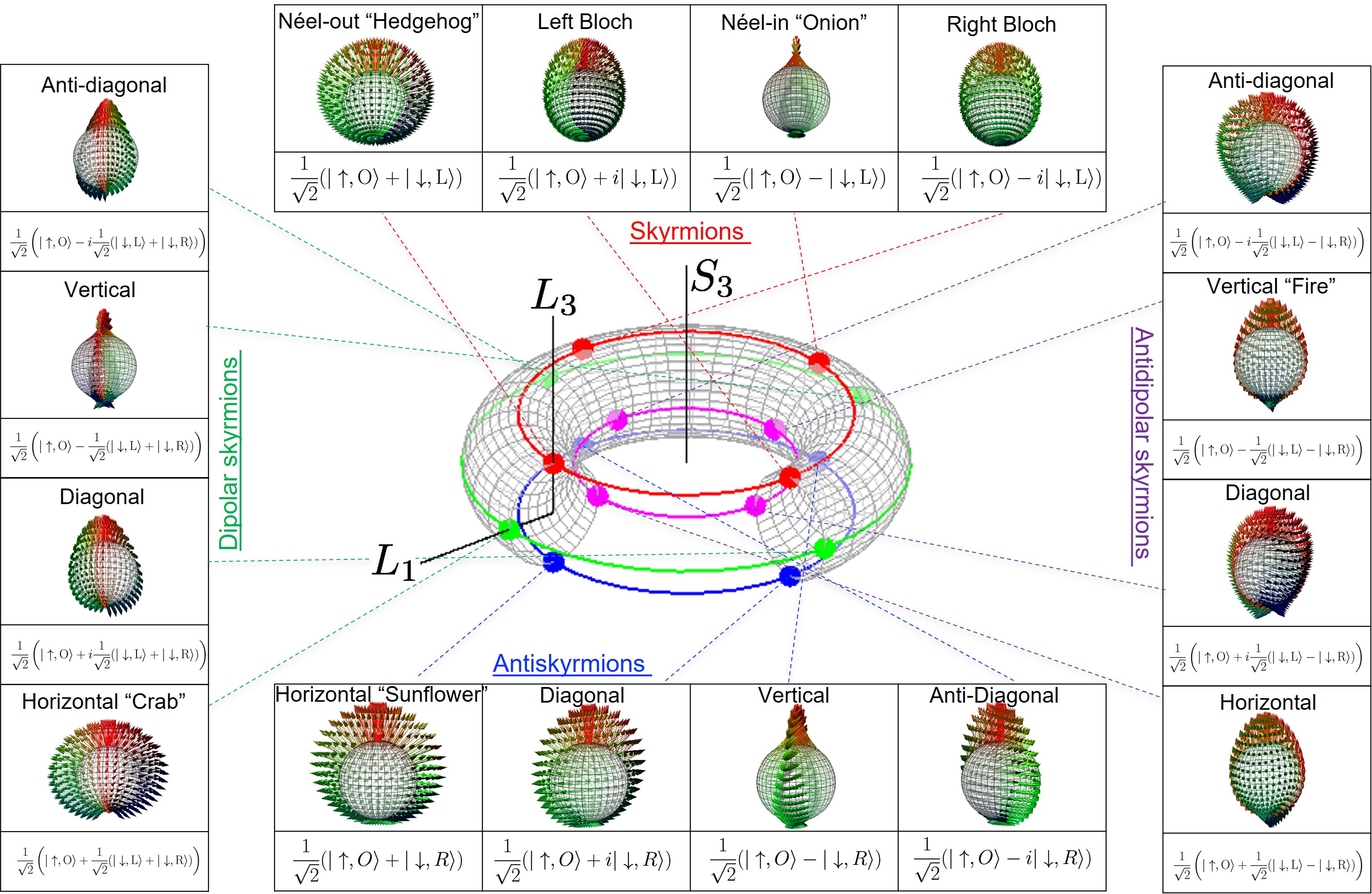}
\caption{
Spin textures on skyrmionic torus ${\mathbb T}^2$.
An arbitrary skrmionic state is described as 
$|\theta_{\rm p}, \phi_{\rm t} \rangle = 
| 3 \rangle / \sqrt{2} +
{\rm e}^{i\phi_{\rm t}}
(\cos(\theta_{\rm p}/2) | 4 \rangle + \sin(\theta_{\rm p}/2)| 5 \rangle)/ \sqrt{2}$, 
where $\theta_{\rm p}$ and $\phi_{\rm t}$ are poloidal and toroidal angles to determine the point $(\theta_{\rm p}, \phi_{\rm t})$ on the torus.
$\theta_{\rm p}$ determines the polar angle on the poloidal circle ${\mathbb T}^1$ to determine amplitudes for left and right vortices, while $\phi_{\rm t}$ determines the azimuthal angle, controlled by the chiral spin angular momentum operator of $\hat{S}_3$.
Calculated spin textures for typical skyrmions are also shown.
}
\end{center}
\end{figure}

\begin{figure}[h]
\begin{center}
\includegraphics[width=16cm]{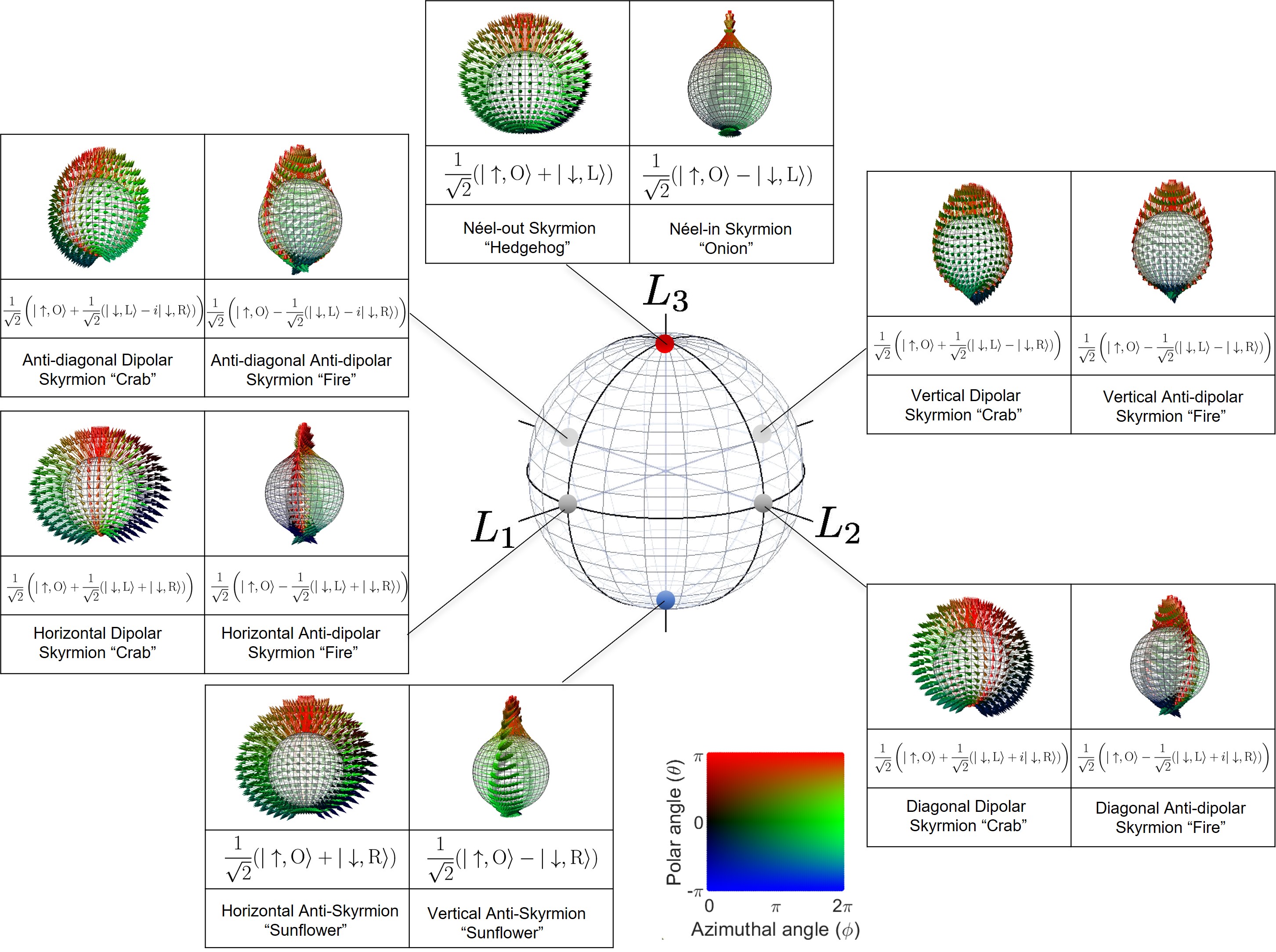}
\caption{
Poincar\'e sphere of optical angular momentum for various skyrmions.
Some skyrmions take the same optical angular momentum, regardless of the orthogonality with different spin textures.
For example, the unitary operation to rotate an N\'eel-out skyrmion along the $L_2$ axis changes the sing upon one rotation to convert it to a N\'eel-in skyrmion, which is orthogonal to the N\'eel-out skyrmion, while both skyrmions give the same expectation values of $(L_1, L_2, L_3)=\hbar N_0/2 (0,0,1)$.
}
\end{center}
\end{figure}

Next, we discuss a continuous unitary transformation from a skyrmion to an antiskyrmion.
Above analysis on Poincar\'e spheres show that a skyrmion is continuously transformed to a pure Guassian state at the north pole of the skrymionic ${\mathbb S}^2$ (Fig. 2a), which also corresponds to the northpole of the antiskyrmionic ${\mathbb S}^2$ (Fig. 2b), such that it can be continuously transformed to an antiskyrmion.
This implies a skyrmion is topologically changed to an antiskyrmion, and in fact, we have a finite overlap between them, as shown by 
$\langle {\rm N\acute{e}el \ out} | {\rm Antiskyrmion} \rangle =1/2$.
However, the topological connection between them is not clear on Poincar\'e spheres.
Our main interest lies in equators (${\mathbb S}^1$) of Poincar\'e spheres to change spin textures for both skyrmions and antiskyrmions, which allow us to consider a skyrmionic torus of ${\mathbb T}^2  \cong {\mathbb S}^1 \times {\mathbb S}^1$, as proposed by Shen {\it et. al} \cite{Shen22b}.
According to our consideration on the SU(6) symmetry, we can identify this non-trivial topological feature actually exists in space spanned by spin angular momentum of $S_3$ together with OAM of $(L_1, L_3)$, as shown in Fig. 9.
Here, we consider a superposition state between a skyrmion and an antiskyrmion, such that the Hilbert space to have at least three orthogonal states are required, which are $|3 \rangle$, $|4 \rangle$, and $|5 \rangle$ in our example shown in Fig. 9, but the other trio ($|1 \rangle$, $|2 \rangle$, and $|6 \rangle$) is also considered by a mirror reflection. 

For the skyrmionic torus \cite{Shen22b}, the state is described by an SU(3) state,  
$|\theta_{\rm p}, \phi_{\rm t} \rangle = 
| 3 \rangle / \sqrt{2} +
{\rm e}^{i\phi_{\rm t}}
(\cos(\theta_{\rm p}/2) | 4 \rangle + \sin(\theta_{\rm p}/2)| 5 \rangle)/ \sqrt{2}$, 
where $\theta_{\rm p}$ is the poloidal angle to change OAM by $\hat{L}_2$ and $\phi_{\rm t}$ is the toroidal angle to change spin by $\hat{S}_3$.
In addition to standard skyrmions and antiskyrmisons, we identify unique spin texutures for dipolar and antidipolar skyrmions (Fig. 9).
These are coming from photonic dipoles, made by superposition states between left and right vortices \cite{Padgett99,Milione11,Naidoo16,Rosales18,Forbes19,Shen19,Shen20b,Shen21,Angelsky21,Yang21,Andrews21,Forbes21,Ma21,Shen22,Nape23}. 
Dipolar skyrmions are made by the superposition states between $|3 \rangle$ and the horizontal dipole of $(|4 \rangle + |5 \rangle)/\sqrt{2}$, with variable phase factor of ${\rm e}^{i\phi_{\rm t}}$.
The spin texture of the dipolar skyrmion is reminiscent of a "Crab", and their face is rotated upon changing $\phi_{\rm t}$.
The skyrmions located at the opposite side along the toroidal direction are orthogonal to each other, similar to polarisation states, such that horizontal and vertical dipolar skyrmions are orthogonal.
Similarly, the antidipolar skyrmion is made by the vertical dipole of $(|4 \rangle - |5 \rangle)/\sqrt{2}$ and $|3 \rangle$.
The spin texture of the antidipolar skyrmion is reminiscent of a "Fire", which also rotates on $\phi_{\rm t}$.

We can change the rotation along the poloidal direction by using $\hat{L}_2$, such that a skyrmion can be continuously changed to an antiskyrmion through the path along the troidal direction. 
The expectation values on the magnitude of OAM determines the radius of the poloidal circle to be $L_{\rm p}=\sqrt{L_1^2+L_3^2}=\hbar N_0/2$, where the factor of two was coming from the fact that a standard Gaussian mode of $|3 \rangle$ does not carry OAM.
It is also noted that states within the poloidal circle under the fixed $\phi_{\rm t}$ are not orthogonal to each other, and the amount of the overlap between $|4 \rangle $ and $|5 \rangle $ can be monitored by the value of $L_3$. 

These photonic skyrmions are also considered in a Poincar\'e sphere for OAM, as shown in Fig. 10.
Here, it is important to recognise SU(2) is double coverage for SO(3), known as ${\rm SU}(2)\cong {\rm SO}(3)/{\mathbb Z}_2$  \cite{Georgi99,Fulton04,Sakurai20,Nakahara03}, such that one rotation over the Poincar\'e sphere changes the phase of the state.
This affects our skyrmionic states, and states with different spin textures can give the same expectation values for OAM (Fig. 10).
For example, N\'eel-out and N\'eel-in skyrmions both give the same expectation values of $(L_1, L_2, L_3)=\hbar N_0/2 (0,0,1)$, while the spin texutres are different.
This is unusual to a standard Poincar'e sphere to describe polarisation, since two orthogonal states usually occupy different points on the Poincar'e sphere.
This clearly shows that one Poincar\'e sphere for OAM is not enough to characterise a skyrmion. 
The orthogonality of these states were confirmed by our skyrmionic Poincar\'e sphere (Fig. 2a), such that we need to monitor several order parameters \cite{Zhang97} to identify the spin textures.
For SU(6) states, in general, we found fifteen Poinca\'e spheres and thirty-five order parameters, which are over-complete to identify a state, composed of ${\mathbb C}^6$, which requires only ${\mathbb R}^{12}$.
The skyrmionic torus \cite{Shen22b} is one of the best ways to clarify the spin texture of skyrmions.
It clearly shows that the topological significance of various skymions and we can continuously transform skyrmions on the torus by using the SU(6) transformation.

\clearpage
\subsection*{Outlook towards future applications}
Finally, we discuss potential applications of our SU(6) theory for photonic skyrmions.
One of the most promising applications would be an optical data storage in silica glass \cite{Sakakura20,Wang24}.
The advanced laser writing technologies allowed researchers to memorise the variable retardance, depending on the pulse duration and the number of pulses \cite{Sakakura20,Wang24}, and the vector vortex beam was successfully observed upon read-out \cite{Sakakura20}.
This means that the local spin texture of photons can be recorded in silica glass, which will be stable for ages \cite{Wang24}.
One can control the chiralities of the spin textures by manipulating the phases and the amplitudes of the laser writing, and the huge Hilbert space of SU(6) would be useful for recording more information in the form of a photonic skyrmion \cite{Ranada89,Beckley10,Beckley12,Tsesses18,Gao20,Lin21,Lei21,Zhu24,Shen22b,Parmee22,McWilliam23,Shen23,Cisowski23,Lin24} 
 or other spin textures such as photonic Bengal cats to form singlet and triplet states \cite{Saito23m}.

Another important application would be fibre optic data transmission for optical communication \cite{Ma21,Liu22}.
Various OAM modes with higher-order topological charges were successfully transmitted and the data transmission as large as 1-Pbps was achieved \cite{Ma21,Liu22}.
We can envisage that we could expand the band width even larger by transmitting photonic skrmions.
However, the effective refractive indices would be different between the fundamental Gaussian mode and the higher-order modes with OAM.
The difference induces an effective field for the SU(6) state, which would be described by the generator of $\hat{\lambda}_8$ for OAM.
This works as an rotator for OAM, which rotates the azimuthal phase on the skyrmionic or antiskyrmionic Poincar\'e spheres, depending on the chirality of OAM.
The difference of the refractive indices also indues the delay of the OAM mode against the fundamental Gaussian mode, which effectively split a skyrmion to two pulses.
Obviously, this will not be ideal for long-haul optical communication, but it might be useful to analyse a photonic skyrmion to identify the OAM state similar to the polarimetry for polarisation states.
Silicon photonic platform will be expected to allow large-scale photonic integrated circuits to generate and detect photonic skrymions \cite{Lin21,Lin24}.

\section*{Conclusions}
We have explored a close relationship between SU(6) symmetry of coherent photons and photonic skyrmions \cite{Ranada89,Beckley10,Beckley12,Tsesses18,Gao20,Lin21,Lei21,Zhu24,Shen22b,
Parmee22,McWilliam23,Shen23,Cisowski23,Lin24}.
We have shown coherent photons has SU(6) symmetry for spin and orbital angular momentum both theoretically and experimentally, and found $\frak{su}$(2) Lie algebra for skyrmions and antiskyrmions.
The expectation values of the generators of rotation are shown on Poincar'e sphere to characterise various skyrmions with unique spin textures upon changing amplitudes and phases by our experimental set-up, named Poincar'e rotator \cite{Saito23k}.
This allows us to consider continuous deformation from a skyrmion to an antiskyrmion by unitary transformation, which was described by a skyrmionic torus \cite{Shen22b}.
The existence of the path to exchange a skyrmion to an antiskyrmion means they are not topologically stable, if there exists an effective field to allow the helical distribution of the local refractive indices.
The standard graded-indexed or step-index fibres have no such helical distribution, such that the exchange would be unlikely to be observed upon the fibre transmission.
On the other hand, if we intentionally apply appropriate SU(6) operations using Poincar'e rotator, it is possible to transform a skyrmion to an antiskyrmion.
We believe a photonic skyrmion serves as a valuable platform to examine various theoretical concepts such as topological structure of coherent photons and associated symmetries in a simple bench-top experiment.

\clearpage
\section*{Methods}

\subsection*{{SU}(6) symmetry of coherent photons}
A single photon with spin ($\sigma=\uparrow , \downarrow$) and OAM ($m$) is described by a state, $|\sigma, m \rangle = |\sigma \rangle_{\rm S} \otimes |m \rangle_{\rm O} = |\sigma \rangle_{\rm S} |m \rangle_{\rm O}$, given by a direct product of states for spin, $|\sigma \rangle_{\rm S}$, and OAM, $ |m \rangle_{\rm O}$.
The total state of $|\sigma, m \rangle$ is characterised by a nonseparability \cite{Spreeuw98,Milione15,McLaren15,Shen22,Shen22b} of spin and OAM, which means that we cannot consider an OAM state without spin or {\it vice versa}.
Consequently, states must be orthogonal to each other, if one of spin or OAM is orthogonal, which can be conveniently represented by the Bose-Einstein commutation relationships \cite{Sakurai20,Georgi99,Weinberg05}, 
$ [ \hat{a}_{\sigma,m} , \hat{a}_{\sigma^{\prime},m^{\prime}}^{\dagger} ] = \delta_{m,m^{\prime}} \delta_{\sigma,\sigma^{\prime}}$ and 
$ [ \hat{a}_{\sigma,m} , \hat{a}_{\sigma^{\prime},m^{\prime}}] =
  [ \hat{a}_{\sigma,m}^{\dagger}  , \hat{a}_{\sigma^{\prime},m^{\prime}}^{\dagger} ] = 0$
, where $\hat{a}_{\sigma,m}^{\dagger}$ and $\hat{a}_{\sigma,m}$ are creation and annihilation operators for a photon and $\delta$ is the Kronecker delta.
We consider an arbitrary polarisation state, given by SU(2) symmetry \cite{Stokes51,Poincare92,Glauber63,Max99}, while OAM states are limited to be $m=0, \pm1$ for a standard Gaussian mode without a vortex (O), together with left (L) and right (R) twisted states, whose superposition states are represented by SU(3) symmetry \cite{Gell-Mann61,Ne'eman61,Gell-Mann64,Georgi99,Fulton04}.
Therefore, we have six orthogonal states, represented by SU(6) symmetry \cite{Georgi99,Weinberg05,Fulton04}, and we label the basis state by an integer, $i=( \sigma, m ) \in \{ 1, \cdots, 6 \} $.
In principle, we can extend OAM to have an arbitrary value of $N \in {\mathbb Z}$, using SU($N$) symmetry, but we restrict our experimental system to have only three modes for OAM, while we consider an arbitrary superposition state among these six states.
After relabelling summarised in Table 1, the commutation relationship is simplified to be $ [ \hat{a}_{i} , \hat{a}_{j}^{\dagger} ] = \delta_{i,j}$. 

\begin{table}[h]
\caption{\label{Table-1}
SU(6) states with spin ($\sigma$) and orbital angular momentum (OAM), labelled by $i=1, \cdots, 6$.
Spin up ($\uparrow$) and down ($\downarrow$) correspond to left and right circularly polarised states.
For OAM, we consider left (L) and right (R) twisted states with topological charges of $m=\pm1$ and a Gaussian state (O) without an optical vortex.
}
\begin{center}
\begin{tabular}{ccc}
\hline
State & Spin  & OAM \\
$i$ & $\sigma$ & $m$ \\
\hline
$|1\rangle = |\uparrow,     {\rm L} \rangle$ & 1 & 1\\
$|2\rangle = |\uparrow,     {\rm R} \rangle$ & 1 & -1\\
$|3\rangle = |\uparrow,     {\rm O} \rangle$ & 1 & 0\\
$|4\rangle = |\downarrow, {\rm L} \rangle$ & -1 & 1\\
$|5\rangle = |\downarrow, {\rm R} \rangle$ & -1 & -1\\
$|6\rangle = |\downarrow, {\rm O} \rangle$ & -1 & 0\\
\hline
\end{tabular}
\end{center}
\end{table}

We consider generators of rotations to realise superposition among these six orthogonal states, which are known for $\frak{su}(6)$ Lie algebra \cite{Gell-Mann61,Ne'eman61,Gell-Mann64,Georgi99,Fulton04}. 
We need to consider $6 \times 6$ matrices, which can be represented by thirty-five generators of rotations ($\frak{\hat{a}}_n$ ($n=1, \cdots, 35$)), while one degree of freedom disappeared due to the traceless condition, corresponding to the determinant of unity for SU(6) Lie group \cite{Gell-Mann61,Ne'eman61,Gell-Mann64,Georgi99,Fulton04}.
Three of them are given by $\bm{\hat{\sigma}} \otimes {\bf \hat{1}}$ for spin, using Pauli matrices of $\bm{\hat{\sigma}}=(\hat{\sigma}_1,\hat{\sigma}_2,\hat{\sigma}_3)$ and the identity operator of $ \hat{\bf 1}$, while eight generators of rotations for OAM are given by  ${\bf \hat{1}} \otimes \bm{\hat{\lambda}}$, where $\bm{\hat{\lambda}} = (\hat{\lambda}_1, \cdots, \hat{\lambda}_8)$ is Gell-Mann matrices for $\frak{su}(3)$ Lie algebra \cite{Gell-Mann61,Ne'eman61,Gell-Mann64,Georgi99,Fulton04}.
These generators of rotations will rotate spin and OAM independently, while the rest of generators $\bm{\hat{\sigma}} \otimes \bm{\hat{\lambda}}$ will rotate them simultaneously \cite{Gell-Mann61,Ne'eman61,Gell-Mann64,Georgi99,Fulton04}.
Overall, we have ${\bf 3} \oplus {\bf 8} \oplus {\bf 24}={\bf 35}$ generators of rotations to cover the $\frak{su}(6)$ Lie algebra \cite{Gell-Mann61,Ne'eman61,Gell-Mann64,Georgi99,Fulton04}. 
The commutation relationship is given by $[ \frak{\hat{a}}_l, \frak{\hat{a}}_m] = 2 i \sum_{n} g_{lmn} \frak{\hat{a}}_n$ for $l,m,n=1,\cdots, 35$, using the structure constant \cite{Haber21,Bossion21} of $g_{lmn}      
=
-
i
\operatorname{tr} 
\left [
[\frak{\hat{a}}_{l}, \frak{\hat{a}}_{m}] \frak{\hat{a}}_{n}
\right ]
/4
$.
The factor of two is included in front of $g_{lmn}$ in the commutation relationship to compensate for the difference in angles between the Hilbert space and angular momentum space.
For example, horizontal and vertical polarisation states are orthogonal for complex electric fields with the angle of $90^{\circ}$, while they are located at the opposite side of the Poincar\'e sphere with the angle of $180^{\circ}$. 

For many-body states, generalised angular momentum operators are defined by
\begin{eqnarray}
\hat{A}_n
=
\hbar
\sum_{i,j=1}^6
\hat{a}_{i}^{\dagger}
\left(
\frak{\hat{a}}_n
\right)_{ij}
\hat{a}_{j},
\end{eqnarray}
for $n=1, \cdots, 35$, and an SU(6) coherent state is given by
\begin{eqnarray}
\lvert \alpha_{1} , \alpha_{2} , \alpha_{3} , \alpha_{4} , \alpha_{5} , \alpha_{6} \rangle
=
\prod_{i=1}^{6}
{\rm e}^{-\frac{\lvert \alpha_{i} \rvert^2}{2}}
{\rm e}^{ \hat{a}_{i}^{\dagger} \alpha_{i} }
\lvert 0 \rangle, 
\end{eqnarray}
where ${\bm \alpha}=(\alpha_{1} , \alpha_{2} , \alpha_{3} , \alpha_{4} , \alpha_{5} , \alpha_{6} ) \in {\mathbb C}^6$ stands for the SU(6) state.
Generalised angular momentum, ${\bm A}=(A_1, \cdots, A_{35})$, is an expectation value, obtained by 
\begin{eqnarray}
A_n
=
\langle \hat{A}_n \rangle
=
\hbar 
{\bm \alpha}^{*}
\frak{\hat{a}}_n
{\bm \alpha},
\end{eqnarray}
as if it is an expectation value for a single photon.
This is coming from the Bose-Einstein condensation nature of coherent photons to occupy a single quantum state described by ${\bm \alpha}$ \cite{Jones41,Fano54,Glauber63,Spreeuw98,Saito23k}.

Generalised angular momentum operators are generators of rotation in this SU(6) Hilbert space.
This can be confirmed by the exponential map from $\frak{su}$(6) Lie algebra to SU(6) Lie group by the unitary transformation along the $l$-th axis
\begin{eqnarray}
\hat{D}_{l}
(\delta \phi)
=
\exp
\left(
-
\frac{i}{\hbar} 
\hat{A}_l
\frac{\delta \phi}{2}
\right)
,
\end{eqnarray} 
which preserves the form of the coherent state \cite{Glauber63,Saito23k}, while the SU(6) state is transferred to be
\begin{eqnarray}
{\bm \alpha}^{\prime}
=
\hat{\mathcal D}_{l}
(\delta \phi)
{\bm \alpha}, 
\end{eqnarray} 
where ${\bm \alpha}$ is transposed to be a column vector, and the SU(6) operator is given by
 the exponential map, 
\begin{eqnarray}
\hat{\mathcal D}_{l}
(\delta \phi)
=
\exp
\left(
-
i
\frak{\hat{a}}_l
\frac{\delta \phi}{2}
\right) 
.
\end{eqnarray} 
The generalised angular momentum after this unitary transformation is given by the quantum-classical correspondence \cite{Saito23k},  
\begin{eqnarray}
A_m^{\prime}
=
\sum_{n=1}^{35}
\left(
{\rm e}^{\hat{G}_{l} \delta \phi}
\right)_{mn}
A_n,
\end{eqnarray} 
between the change in the SU($N$) wavefunction and its expectation value as the SO($N^2$ -1) angular momentum, 
where $\hat{G}_{l}$ is the adjoint operator, whose matrix element is given by the structure constant of $g_{lmn}$ as $(\hat{G}_{l})_{mn}=-g_{lmn}$, which satisfies the commutation relationship of $[ \hat{G}_{l}, \hat{G}_{m}] = \sum_{n=1}^{35} g_{lmn} \hat{G}_{n}$.
This means that the generalised angular momentum is rotated on the hypersphere of ${\mathbb S}^{35}$ by the special orthogonal group of the degree thirty-five, SO(35).
In fact, we could prove \cite{Saito23k} that the norm is conserved upon the rotation, and the radius of the hypersphere becomes $A_0=\hbar N_0 \sqrt{5/3}$.

More generally, the rotational axis could be pointing to an arbitrary direction of $\hat{\bf n}=(n_1,\cdots, n_{35})$ in SO(35) space, whose unitary transformation becomes
\begin{eqnarray}
\hat{D}_{\hat{\bf n}}
(\delta \phi)
=
\exp
\left(
-
\frac{i}{\hbar} 
\hat{\bf A} \cdot \hat{\bf n}
\frac{\delta \phi}{2}
\right)
,
\end{eqnarray} 
where $\hat{\bf A} = (\hat{A}_1,\cdots, \hat{A}_{35})$.
Lie algebra is vector space \cite{Gell-Mann61,Ne'eman61,Gell-Mann64,Georgi99,Fulton04}, such that the vectorial sum rule is applicable and $\hat{\bf A} \cdot \hat{\bf n}$ also works as a generator of rotation.
It is a straightforward calculation using the Campbell-Baker-Hausdorff formula to confirm the hyperspherical relationship \cite{Saito23k}  
\begin{eqnarray}
{\bm A}^{\prime}
=
\left(
{\rm e}^{\hat{\bf G} \cdot \hat{\bf n} \delta \phi}
\right)
{\bm A},
\end{eqnarray} 
where the standard matrix product calculus should be applied using the transpose of ${\bm A}$, and $\hat{\bf G} = (\hat{G}_1,\cdots, \hat{G}_{35})$ is angular momentum operator in the adjoint representation.
This formula is applicable to an arbitrary value of $N$ for an SU($N$) state \cite{Saito23m}.

\subsection*{$\frak{su}$(2) subalgebra for skyrmions and antiskyrmions}
We consider a map from an SU(6) state, described by ${\mathbb C}^6$, to associated expectation values of generalised angular momentum on ${\mathbb S}^{35}$.
In general, we can realise an arbitrary superposition state with varying amplitudes and phases among six orthogonal states.
If we focus on mixing of two orthogonal states, they are described by the $\frak{su}$(2) subalgebra, whose expectation values are represented on an Poincar\'e sphere of ${\mathbb S}^{2}$.
We have $6\cdot5/2=15$ ways of selecting two states among six states, such that we can envisage fifteen Poincar\'e spheres.
Three of them are standard Poincar\'e spheres on polarisation for each pure OAM state (superposition states between 
$|1 \rangle$ and $|4 \rangle$, $|2 \rangle$ and $|5 \rangle$, or, $|3 \rangle$ and $|6 \rangle$); 
six of them are Poincar\'e spheres for OAM under fixed polarisation 
(
$|1 \rangle$ and $|2 \rangle$, $|1 \rangle$ and $|3 \rangle$, $|2 \rangle$ and $|3 \rangle$, $|4 \rangle$ and $|5 \rangle$, $|4 \rangle$ and $|6 \rangle$, or, $|5 \rangle$ and $|6 \rangle$);
two of them are singlet-triplet ($|2 \rangle$ and $|4 \rangle$) and triplet-triplet coupling ($|1 \rangle$ and $|5 \rangle$).
We have already examined these Poincar\'e spheres previously to confirm the symmetry and a classical entanglement \cite{Glauber63,Spreeuw98,Sasada03,Forbes21,Nape23,Ma21,Rosen22,Shen21,Shen22,Cisowski22,Lopes19,Saito23k}.
The rest of four Poincar\'e spheres are appropriate to examine the coupling among optical modes with and without OAM while chirality of spin is opposite to observe  
skyrmions 
(
$|3 \rangle$ and $|4 \rangle$, or, $|2 \rangle$ and $|6 \rangle$) and 
antiskyrmions 
(
$|3 \rangle$ and $|5 \rangle$, or, $|1 \rangle$ and $|6 \rangle$).
Among these four combinations, the combination between $|3 \rangle$ and $|4 \rangle$ ($|3 \rangle$ and $|5 \rangle$) and $|2 \rangle$ and $|6 \rangle$ ($|1 \rangle$ and $|6 \rangle$) are just mirror symmetric both for spin and OAM upon reflection, and here, we just focus only on the combination of $|3 \rangle$ and $|4 \rangle$ ($|3 \rangle$ and $|5 \rangle$) for skyrmions (antiskyrmions).

Lie algebra is vector space, such that we can define another generators of rotation by summing up $\frak{su}$(6) generators with appropriate factors to change the axis for rotation.
For skyrmions, it is useful to define skyrmionic generators of rotation
\begin{eqnarray}
\hat{\mathcal S}_1
&=&
\frac{1}{2}
\left(
\sigma_1 \otimes \lambda_4
+
\sigma_2 \otimes \lambda_5
\right) 
=\hat{\frak b}_{13}
\\
\hat{\mathcal S}_2
&=&
\frac{1}{2}
\left(
-
\sigma_1 \otimes \lambda_5
+
\sigma_2 \otimes \lambda_4
\right) 
=\hat{\frak b}_{14}
\\
\hat{\mathcal S}_3
&=&
\sigma_3 \otimes 
\left(
\frac{1}{3} {\bm 1}
+
\frac{1}{4} \lambda_3
-
\frac{\sqrt{3}}{12} \lambda_8
\right)
-
{\bm 1} \otimes 
\left(
\frac{1}{4}
\lambda_3
+
\frac{\sqrt{3}}{4}
\lambda_8
\right) 
=
-\frac{\sqrt{3}}{3}
\hat{\frak b}_{8}
+\frac{\sqrt{6}}{3}
\hat{\frak b}_{15}, 
\end{eqnarray} 
which work similarly to Pauli matrices of $(\hat{\sigma}_1, \hat{\sigma}_2, \hat{\sigma}_3)$ for $|3 \rangle$ and $|4 \rangle$ states for $\frak{su}$(2) subalgebra, and $\hat{\frak b}_{i}$ ($i=1,\cdots,35$) is a generator of rotation for $\frak{su}$(6) algebra in a Cartan-Dynkin formalism \cite{Georgi99}.
Similarly, for antiskyrmions, we define antiskyrmionic generators of rotation 
\begin{eqnarray}
\hat{\mathcal A}_1
&=&
\frac{1}{2}
\left(
\sigma_1 \otimes \lambda_6
+
\sigma_2 \otimes \lambda_7
\right) 
=\hat{\frak b}_{20}
\\
\hat{\mathcal A}_2
&=&
\frac{1}{2}
\left(
-
\sigma_1 \otimes \lambda_7
+
\sigma_2 \otimes \lambda_6
\right) 
=\hat{\frak b}_{21}
\\
\hat{\mathcal A}_3
&=&
\sigma_3 \otimes 
\left(
\frac{1}{3} {\bm 1}
-
\frac{1}{4} \lambda_3
-
\frac{\sqrt{3}}{12} \lambda_8
\right)
+
{\bm 1} \otimes 
\left(
\frac{1}{4}
\lambda_3
-
\frac{\sqrt{3}}{4}
\lambda_8
\right) 
=-\frac{\sqrt{3}}{3}
\hat{\frak b}_{8}
+\frac{\sqrt{6}}{12}
\hat{\frak b}_{15}
+
\frac{\sqrt{10}}{4}
\hat{\frak b}_{24}
\end{eqnarray} 
to define another $\frak{su}$(2) subalgebra, describing for the coupling between $|3 \rangle$ and $|5 \rangle$.
They are described by $2 \times 2$ Pauli matrices for the Hilbert space, spanned by $|3 \rangle$ and $|4 \rangle$ ($|3 \rangle$ and $|5 \rangle$), for skyrmions and antiskyrmions, respectively.
For many-body coherent states, we define skyrmionic and antiskyrmionic SU(2) generators of rotation 
\begin{eqnarray}
\hat{\frak S}_{j}
&=&
\hbar
\left (
  \begin{array}{cc}
\hat{a}_{3}^{\dagger} & \hat{a}_{4}^{\dagger}
  \end{array}
\right)  
\hat{\mathcal S}_{j}
\left (
  \begin{array}{c}
\hat{a}_{3} \\
\hat{a}_{4}  
  \end{array}
\right) 
 \\
\hat{\frak A}_{j}
&=&
\hbar
\left (
  \begin{array}{cc}
\hat{a}_{3}^{\dagger} & \hat{a}_{5}^{\dagger}
  \end{array}
\right)  
\hat{\mathcal A}_{j}
\left (
  \begin{array}{c}
\hat{a}_{3} \\
\hat{a}_{5}  
  \end{array}
\right) 
, 
\end{eqnarray}
for $j=1,2,3$, whose expectation values are $({\frak S}_{1},{\frak S}_{2},{\frak S}_{3})$ and $({\frak A}_{1},{\frak A}_{2},{\frak A}_{3})$, respectively.
The operators $\hat{\frak S}_{j}$ and $\hat{\frak A}_{j}$ are made of $\frak{su} (6)$ operators, but they are essentially equivalent to  $\frak{su} (2)$ operators for skyrmions and anitiskyrmions, respectively, such that we can utilise generalised Euler's formulas
\begin{eqnarray}
\exp
\left(
-i
\hat{\mathcal S}_{j}
\frac{\delta \phi}{2}
\right) 
&=&
{\bf 1}
\cos
\left(
\frac{\delta \phi}{2}
\right) 
-
i
\hat{\mathcal S}_{j}
\sin 
\left(
\frac{\delta \phi}{2}
\right) 
 \\
\exp
\left(
-i
\hat{\mathcal A}_{j}
\frac{\delta \phi}{2}
\right) 
&=&
{\bf 1}
\cos
\left(
\frac{\delta \phi}{2}
\right) 
-
i
\hat{\mathcal A}_{j}
\sin 
\left(
\frac{\delta \phi}{2}
\right) 
, 
\end{eqnarray} 
which are continuous transformations from the identity operator of ${\bf 1}$.

\subsection*{Experimental setup}
Our experimental setup of a Pincar\'e rotator is shown in Fig. 1. 
We used a Laser Diode (LD), which was operated under constant current of 220mA at the wavelength was 532 nm.
The beam shape was adjusted by a collimator lens (CL) with the focal length of 100mm, followed by a pin hole (PH0) with the diameter of 200 ${\rm \mu}$m.
We used standard optical components to control polarisation such as a polariser (PL), a half-wave-plate (HWP), a quarter-wave-plate (QWP), a polarisation-beam-splitter (PBS), a mirror (M), a non-polarisation-beam-splitter (NPBS), a polarimeter (PM), and a complementary-metal-oxide-semiconductor camera (CMOS).
Here, we have mechanically rotated HWP1 to change amplitudes before PBS, while the series of operations by QSP1, HWP2, HWP3, and QWP2 were employed to control the phase, whose amount was controlled by the rotation angle of HWP3.
QWP3 and QWP4 were used to rotate the polarisation state to be left or right circularly polarised state.
The polarisation state was checked by a PM with optional PL2.
The final image was taken by CMOS with and without PL2.
We took far-field images for six polarisation states of $|{\rm L} \rangle$, $|{\rm R} \rangle$, $|{\rm H} \rangle$, $|{\rm V} \rangle$, $|{\rm D} \rangle$, and $|{\rm A} \rangle$ using waveplates and polarisers to observe the spin texture. 



\clearpage

\bibliography{Saito}

\section*{Acknowledgements}
The author would like to express sincere thanks to Prof I. Tomita for continuous discussions and encouragements. 
This work is supported by JSPS KAKENHI Grant Number JP 18K19958.

\section*{Author contributions statement}
The author confirms being the sole contributor of this work and has approved it for publication.

\section*{Competing interests} 
The author declares no conflict of interest. 



\end{document}